\documentclass[10pt,journal,compsoc]{IEEEtran}
\ifCLASSOPTIONcompsoc
  \usepackage[nocompress]{cite}
\else
  \usepackage{cite}
\fi
\ifCLASSINFOpdf
   \usepackage[pdftex]{graphicx}
\else
   \usepackage[dvips]{graphicx}
\fi
\usepackage{xcolor}

\usepackage{amsmath}
\usepackage{algorithmic}

\usepackage{booktabs} 

\ifCLASSOPTIONcompsoc
 \usepackage[caption=false,font=footnotesize,labelfont=sf,textfont=sf]{subfig}
\else
 \usepackage[caption=false,font=footnotesize]{subfig}
\fi

\usepackage{stfloats}
\ifCLASSOPTIONcaptionsoff
 \usepackage[nomarkers]{endfloat}
\let\MYoriglatexcaption\caption
\renewcommand{\caption}[2][\relax]{\MYoriglatexcaption[#2]{#2}}
\fi
\usepackage{url}

\usepackage{upgreek}
\usepackage{placeins}

\begin{document}
%
\title{Overview of the IBM Neural Computer Architecture}
%
%
%
%

\author{Pritish Narayanan,
        Charles E. Cox,
        Alexis Asseman,
        Nicolas Antoine,
        Harald Huels,
        Winfried W. Wilcke,
        
        and~Ahmet S. Ozcan
\IEEEcompsocitemizethanks{\IEEEcompsocthanksitem The authors, except H. Huels, are with the IBM Research Division, Almaden Research Center, San Jose, CA 95120. H. Huels is with IBM Boeblingen, Germany.\protect\\

E-mail: pnaraya@us.ibm.com}
\thanks{Manuscript revised January 29, 2020.}}

%
%

\markboth{IEEE Transactions, DRAFT MANUSCRIPT, Feb 2020}%
{Shell \MakeLowercase{\textit{et al.}}: Bare Demo of IEEEtran.cls for Computer Society Journals}
%



\IEEEtitleabstractindextext{%
\begin{abstract}
The IBM Neural Computer (INC) is a highly flexible, re-configurable parallel processing system that is intended as a research and development platform for emerging machine intelligence algorithms and computational neuroscience. It consists of hundreds of programmable nodes, primarily based on Xilinx's Field Programmable Gate Array (FPGA) technology. The nodes are interconnected in a scalable 3d mesh topology. We overview INC, emphasizing unique features such as flexibility and scalability both in the types of computations performed and in the available modes of communication, enabling new machine intelligence approaches and learning strategies not well suited to the matrix manipulation/SIMD libraries that GPUs are optimized for. This paper describes the architecture of the machine and applications are to be described in detail elsewhere.
\end{abstract}

}

\maketitle

\IEEEdisplaynontitleabstractindextext

%
\IEEEpeerreviewmaketitle

\IEEEraisesectionheading{\section{Introduction}\label{sec:introduction}}

%
%
%
%
\IEEEPARstart{T}{he} revolution in deep learning over the last decade has been mainly driven by the confluence of two equally important factors -- the generation of large amounts of data, and the availability of GPUs to train large neural networks. Those contributed to accelerated research in the algorithmic domain by significantly decreasing experiment turn-around time. In recent years, as the computational demands for deep learning have increased, especially in consumer facing domains such as image and speech recognition, there is a trend towards more hardware specialization to improve performance and energy-efficiency. For instance, recent hardware approaches \cite{sijstermans:2018,dally:2018,courbariaux:2014,gupta:2015,han:2015,cchen:2018,jouppi:2018} feature techniques such as reduced precision, aggressive compression schemes and customized systolic data paths aimed at accelerating today's DNNs. 

While these efforts are extremely important both from research and commercialization perspectives, the currently available hardware landscape is also somewhat restrictive. This is because GPUs (and their variants) may not lend themselves well to use cases where mini-batching/data parallelism is non-trivial because multiply-accumulate operations are not dominant or when low latency is critical. Therefore, as machine intelligence algorithms continue to evolve, it is unfortunate that promising approaches may be sidelined simply because they do not map well to a GPU, just as backpropagation trailed more conventional machine learning approaches for decades due to the lack of GPUs. One of the prime examples of an algorithm which is not well matched to SIMD architecture is Monte Carlo Tree Search used in the Google Deepmind's AlphaGo system\cite{SilverHuangEtAl16nature}.


This indicates the need for new computer architectures for machine intelligence. However, there is an obvious challenge -- namely, how does one build hardware for algorithms and use cases that may not yet exist? The approach we have taken is to build a hardware system with tremendous flexibility, which will be explored in depth in the rest of the paper. This flexibility extends to the types of algorithms being executed, the portions that need to be off-loaded and accelerated, the model of communication, and the types of parallelism deployed.

\begin{figure*}[!t]
    \centering
	\includegraphics[width=0.8\textwidth]{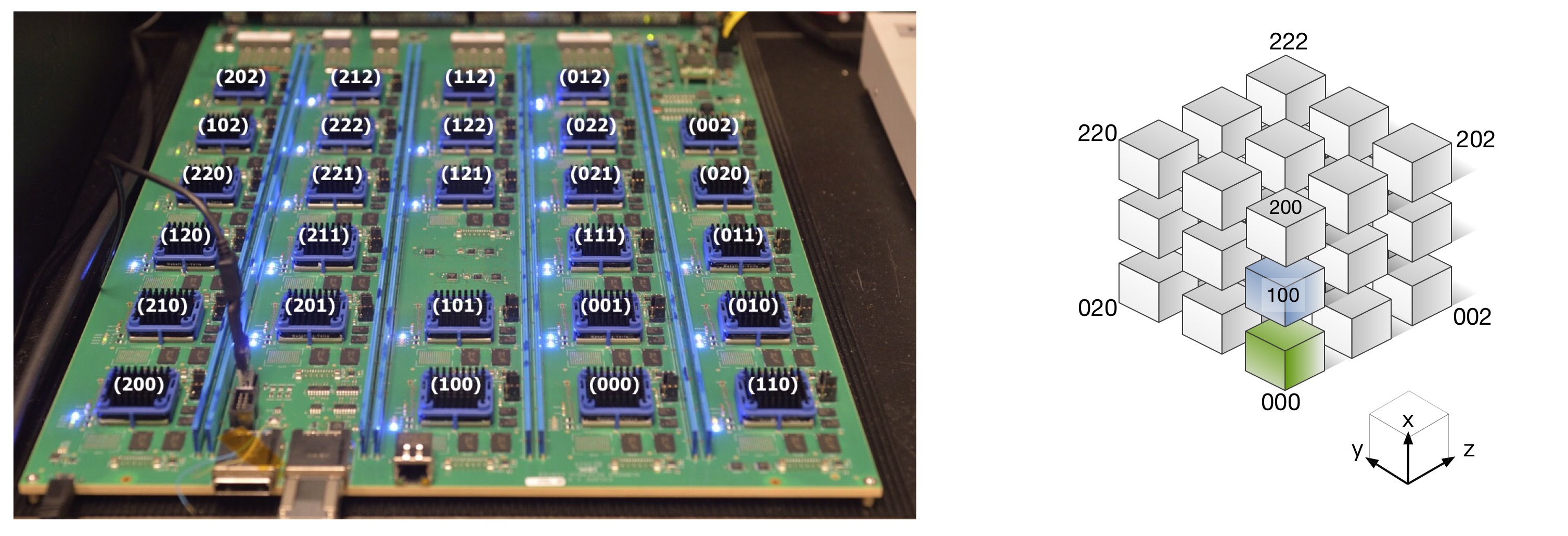}
\caption{3D Mesh Topology of a single INC card, with some node numbers shown. Each node contains its own dedicated Zynq System-on-Chip with an ARM Cortex-A9 and FPGA logic. Node 100 (blue shaded cube) is the gateway node to the external Ethernet. Nodes 000 (green shaded cube) and 200 have PCIe Interfaces to communicate with a host computer.}
	\label{fig:INC_card}
\end{figure*}

The system, named IBM Neural Computer (INC), is fundamentally a large, highly scalable parallel processing system with compute nodes interconnected in a 3D mesh topology. The total number of compute nodes within the existing single-cage system is 432. Each compute node contains a Xilinx Zynq System-on-Chip (an ARM A9 CPU + FPGA logic on the same die) along with 1GB of dedicated RAM \cite{rajagopalan:2011,crockett:2014,zynq7000}. The availability of FPGA resources on every node allows application-specific processor offload, a feature that is not available on any parallel machine of this scale that we are aware of. 

The communication network that realizes the 3D mesh topology is implemented using single-span and multi-span SERDES (Serializer-Deserializer) links connected to the FPGA fabric of the Zynq. It is therefore possible to build tailored hardware network controllers based on the communication mode(s) most suited to the application. The ability to optimize the system performance across application code, middle-ware, system software, and hardware is a key feature of INC.

One may envision that this 3D topology of distributed memory and compute, with the ability to have nodes exchange signals/messages with one another is somewhat reminiscent of works targeting the human brain (most prominently, the SpiNNaker project\cite{furber:2014,painkras:2013}). However, an important distinction is that our goal is more general than computational neuroscience, and extends to machine intelligence in general.

The rest of the paper is organized as follows. Section \ref{sec:xcape9000} begins with an overview of the INC system. Section \ref{sec:comm} discusses inter-node communication mechanisms in INC. Section \ref{sec:diag} contains details on diagnostic and debug capabilities of the INC system, which are crucial in a development environment. Section \ref{sec:conclusion} will conclude the paper.

\begin{figure*}[!t]
    \centering
	\includegraphics[width=0.8\textwidth]{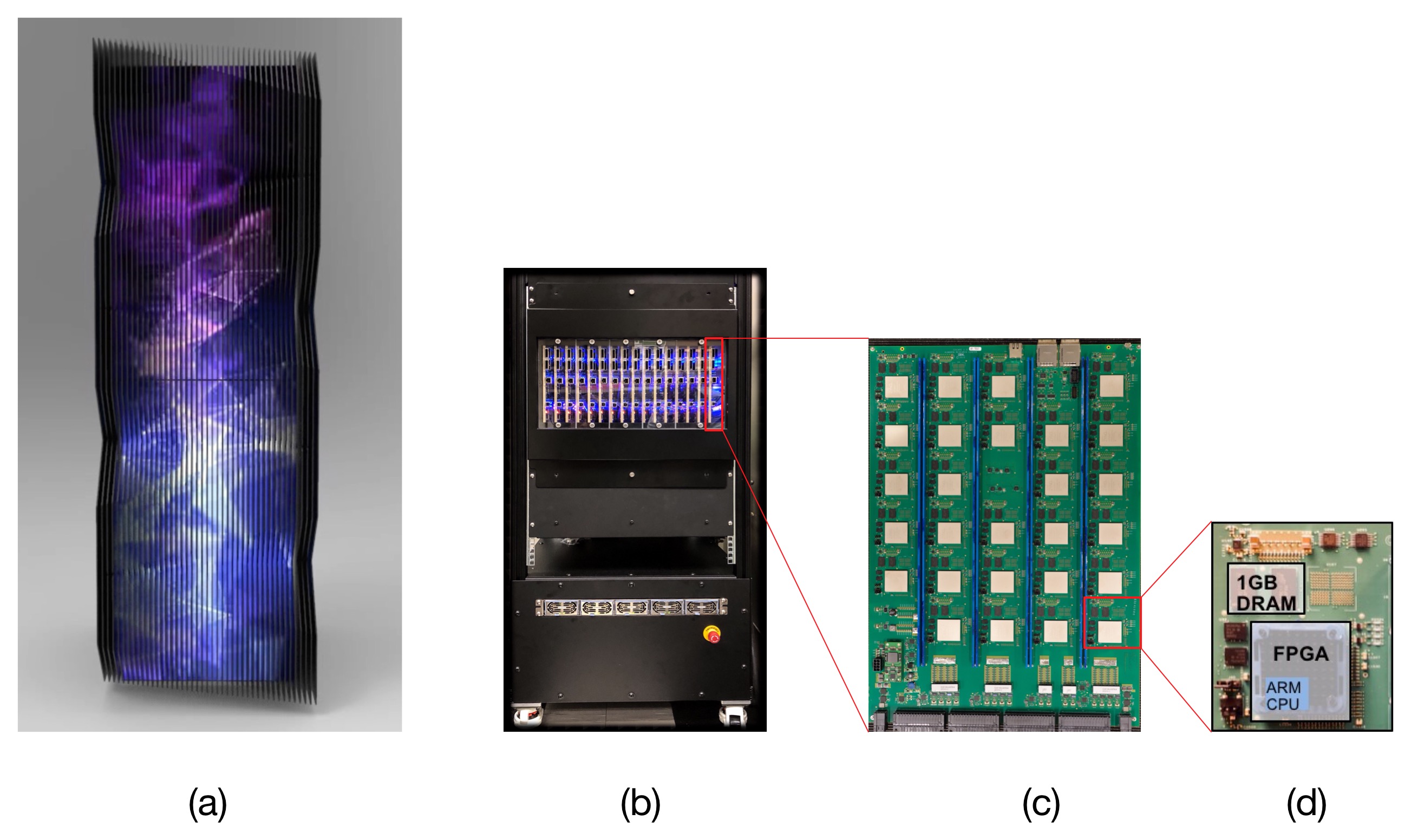}
\caption{Hierarchical organization of the INC system (a) Conceptual design of INC 9000 with 48 cards and  1296 nodes (not yet built), (b) INC 3000 with 16 cards and 432 nodes (operational), (c) Single card with 27 nodes, (d) One node.}
	\label{fig:INC_system}
\end{figure*}


\section{INC Overview}\label{sec:xcape9000}

The INC system is designed primarily to be a development platform for emerging machine intelligence algorithms. It is a parallel processing system with a large number of compute nodes organized in a high bandwidth 3D mesh network. The platform is designed to be highly flexible. Within each node is a Xilinx Zynq system-on-chip, which integrates a dual-core Cortex A9 ARM processor and an FPGA on the same die, allowing the system to be reconfigured on a per node basis. Each node also includes 1GB of dedicated DRAM that can be used as program and data space, and is accessible both from the processor and the FPGA. In an eventual at-scale high-performance learning task, we envision that most of the performance critical steps will be offloaded and optimized on the FPGA, with the ARM only providing auxiliary support -- including initialization, diagnostics, output transfer, etc. 

While INC is a distributed system in that it is composed of distinct processor+memory nodes interconnected by communication links, it has a unique combination of features not available elsewhere. It is not a multi-FPGA `sea of gates' system \cite{krupnova:2004,assad:2012} whose structure would need to be defined by the logic resident on the FPGA. It has a very well defined structure of compute nodes with a well defined communications network. Therefore it does not carry the performance compromise associated with the need to support a fully-generic interconnect. 

It is also different from other distributed systems such as BlueGene \cite{haring:2012}. In addition to the available FPGA offload capability at every node, the communication interfaces are not pre-defined to support a limited set of known use cases. Instead, access to the physical communication links is through the FPGA, and multiple distinct `logical' channels of communication can be established, all utilizing the same underlying SERDES links. In this way, the network interfaces can be designed (and even progressively optimized) to best suit the applications executing on INC. 


\subsection{INC card}

The basic building block of the system is an INC card. Each card contains 27 nodes arranged in a 3$\times$3$\times$3 cube. 

Figure~\ref{fig:INC_card} shows the 3x3x3 topology of an individual card, along with (XYZ) co-ordinates overlaid to indicate the organization of the 3D mesh. The 27 nodes are placed on the card in a way to minimize the connection lengths between logically adjacent nodes. All 27 nodes on a single card are identical except for some important differences. Node (100) includes an Ethernet port, and can act as a gateway connecting an internal Ethernet network implemented on the FPGAs to a conventional external network. Node (000) is a controller node, and includes a 4 lane PCIe 2.0 connection that can be connected to a host PC. It also has a serial connection that can serve as a console during boot time, or be forwarded to the other nodes on the card. Node (200) is also capable of supporting a PCIe interface, should an application need additional bandwidth. 

\subsection{Backplane, Cages and Racks}


In an INC system, individual cards plug into a backplane. Each backplane can support up to 16 cards, and the backplane wiring arranges the 432 nodes of the 16 cards into a 12$\times$12$\times$3 mesh. The backplane and cards are enclosed in an INC card cage (INC 3000 system (Fig.~\ref{fig:INC_system}b)). Connectors on the back side of the backplane allow up to four cages to be connected vertically to build a system of up to 12$\times$12$\times$12 or 1728 nodes (INC 9000 system (Fig.~\ref{fig:INC_system}a)).

The card design supports building an INC system with anywhere from one to 512 cards (13,824 nodes). However, to grow beyond the size of the INC 9000 system, a new backplane design is required.

\subsection{Physical Links}
Each node on a card is connected to its nearest orthogonal neighbors by a single span link composed of two unidirectional serial connections. Each node has six single span links. The nodes on the faces of the cube (i.e. all nodes other than the central (111) node) have single span links that leave the card, and may have nearest neighbors on other cards in the system. In addition to single span links, 6 bi-directional multi-span links allow for more efficient communication in a larger system. Multi-span links connect nodes that are three nodes apart in any one orthogonal direction, and will always begin and terminate on different cards. With a total of 432 links leaving or entering the card, and 1 Giga-byte (GB) per second per link, this amounts to a potential maximum bandwidth of 432 GB per second leaving and entering one card. The bisection bandwidths for the INC 9000 and INC 3000 systems are 864 GB per second and 288 GB per second respectively.

The communications links are pairs of high speed, serial, unidirectional SERDES connections. Each connection only has two wires: the differential data lines. There are no additional lines for handshake signals. The links are controlled by a credit scheme to ensure that overrun errors do not occur and no data is lost. A receiving link sends (via its paired transmit link) a count of how many bytes of data it is willing to receive. A transmitting link will decrement its count as it sends data and never send more data than it holds credits for from the receiver. The receiving side will add to the credit balance as it frees up buffer space. This credit system is implemented entirely in the hardware fabric and does not involve the ARM processor or software. 

\subsection{Packet Routing}

The communication network currently supports directed and broadcast packet routing schemes. Features such as multi-cast or network defect avoidance are being considered at the time of writing, and can be included based on application or hardware needs. 

In a directed routing mode, a packet originating from the processor complex or the FPGA portion of a compute node is routed to a single destination. Both single-span and multi-span links may be used for the routing, and the packet will be delivered with a minimum number of hops. However, a deterministic routing path is not guaranteed, as each node involved may make a routing decision based on which links happen to be idle at that instant. This implies that in-order delivery of packets is not guaranteed\footnote{This does not preclude applications that need in-order delivery, as reordering can be achieved in either FPGA hardware or in software, or a different packet routing scheme can be devised as necessary.}. The packet routing mechanism is implemented entirely on the FPGA fabric, and the ARM processors may only be involved at the source and destination nodes, if at all. 

A broadcast packet radiates out from the source node in all directions and is delivered to every node in the system. Broadcast packets only use the single-span links in the system for simplicity of routing. Depending on which link received a broadcast packet, the receiving node may choose to a) forward to all other links, b) forward to a subset of links, or c) stop forwarding. By choosing the rules for these three scenarios carefully, it is possible to ensure that all nodes in the system receive exactly one copy of the broadcast packet. 

\section{Connectivity and Communication}\label{sec:comm}

Multiple virtual channels can be designed to sit atop the underlying packet router logic described in the previous section  to give the processor and FPGA logic different virtual or logical interfaces to the communication network.  In this section we will review three approaches currently implemented on INC -- Internal Ethernet, Postmaster Direct Memory Access (DMA) and Bridge FIFO. 

\subsection{Internal Ethernet}

\begin{figure*}[!t]
    \centering
	\includegraphics[width=.85\textwidth]{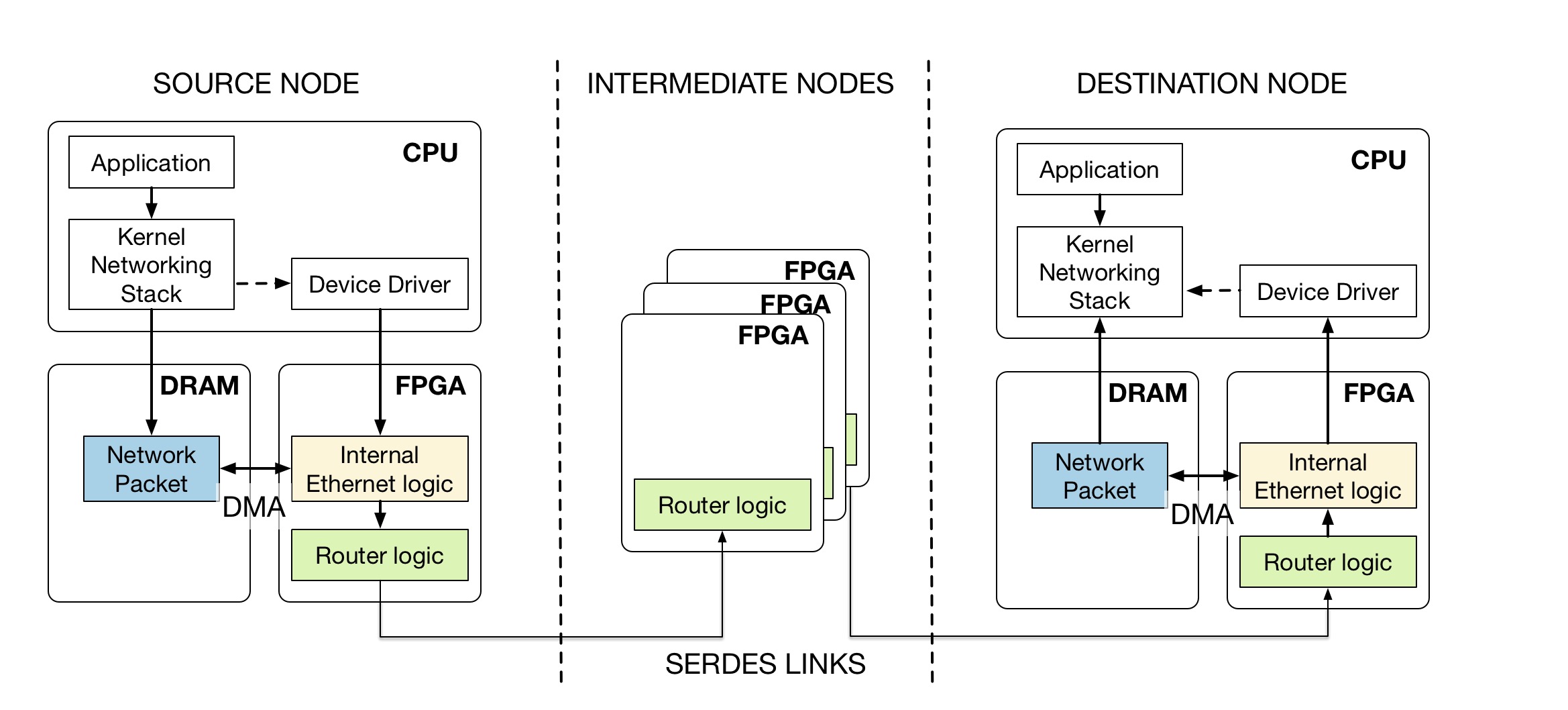}
\caption{Operation of the virtual internal Ethernet interface: Network packets generated at a source node are initially in its DRAM. At the request of the Ethernet Device driver, a DMA transfer is initiated and the packet is transferred to the router logic facing the SERDES links. The packet will traverse through zero or more intermediate nodes without processor interaction before reaching the destination, where the Ethernet device implemented on the FPGA fabric will raise a hardware interrupt, notifying the driver and thereby the kernel of the new packet to be processed.}
	\label{fig:eth_int}
\end{figure*}

One of the virtual interfaces is designed to appear similar to an Ethernet interface\footnote{Note that this is an interface for node to node communication, and is different from the `real' physical Ethernet interface at node (100) which is intended for communicating with the external world.}. While the underlying hardware is quite different from an Ethernet network, this design point is chosen to take advantage of the large amount of standard application software readily available for IP networks such as \texttt{ssh}. A Linux OS and associated device driver running on an ARM processor can then use these applications to communicate with other nodes on the internal network operating as if it were communicating with a real Ethernet device. Similarly, applications that depend on standard parallel software libraries (e.g. Message Passing Interface (MPI) \cite{geist:1996} and its variants) can be easily supported. Using stable, well-established networking applications was also extremely useful during initial debugging of the network hardware and the system software.

The operating mechanism for transmitting packets is conceptualized in Fig.~\ref{fig:eth_int}. During the Transmit Operation, the application passes information to the kernel networking stack, which adds various headers and sends it on to a virtual internal Ethernet interface (ethX).
This interface is owned by the device driver, which manages a set of buffer descriptors that contain information about the size and memory locations of various packets in the DRAM. 
The device driver then informs the hardware about the availability of a packet to be transmitted, by setting a status bit. 
The actual transfer from the DRAM into the FPGA fabric is a DMA operation, using an AXI-HP bus on the Zynq chip. 
Packet receive is conceptually a reverse operation, with the distinction that the device driver has two mechanisms to know of the arrival of a packet on the interface -- one is a hardware interrupt, and the other is a polling mechanism that is far more efficient under high traffic conditions. 
Note that while this description assumes applications running in software as the producers and consumers of the packets, the internal Ethernet can also be accessed from other hardware blocks on the FPGA fabric itself, if at all necessary. 

The availability of this virtual internal Ethernet also makes it straightforward for any node in the system to communicate with the external world using TCP/IP. This is done by using the physical Ethernet port on node (100) and configuring this node as an Ethernet gateway implementing Network Address Translation (NAT) and port forwarding. One immediate and obvious use of this feature is the implementation of an NFS (network file system) service to save application data from each of the nodes (whose file systems are implemented on the DRAM and are therefore volatile) to a non-volatile external storage medium. 

\subsection{Postmaster DMA}

\begin{figure*}[!t]
    \centering
	\includegraphics[width=.85\textwidth]{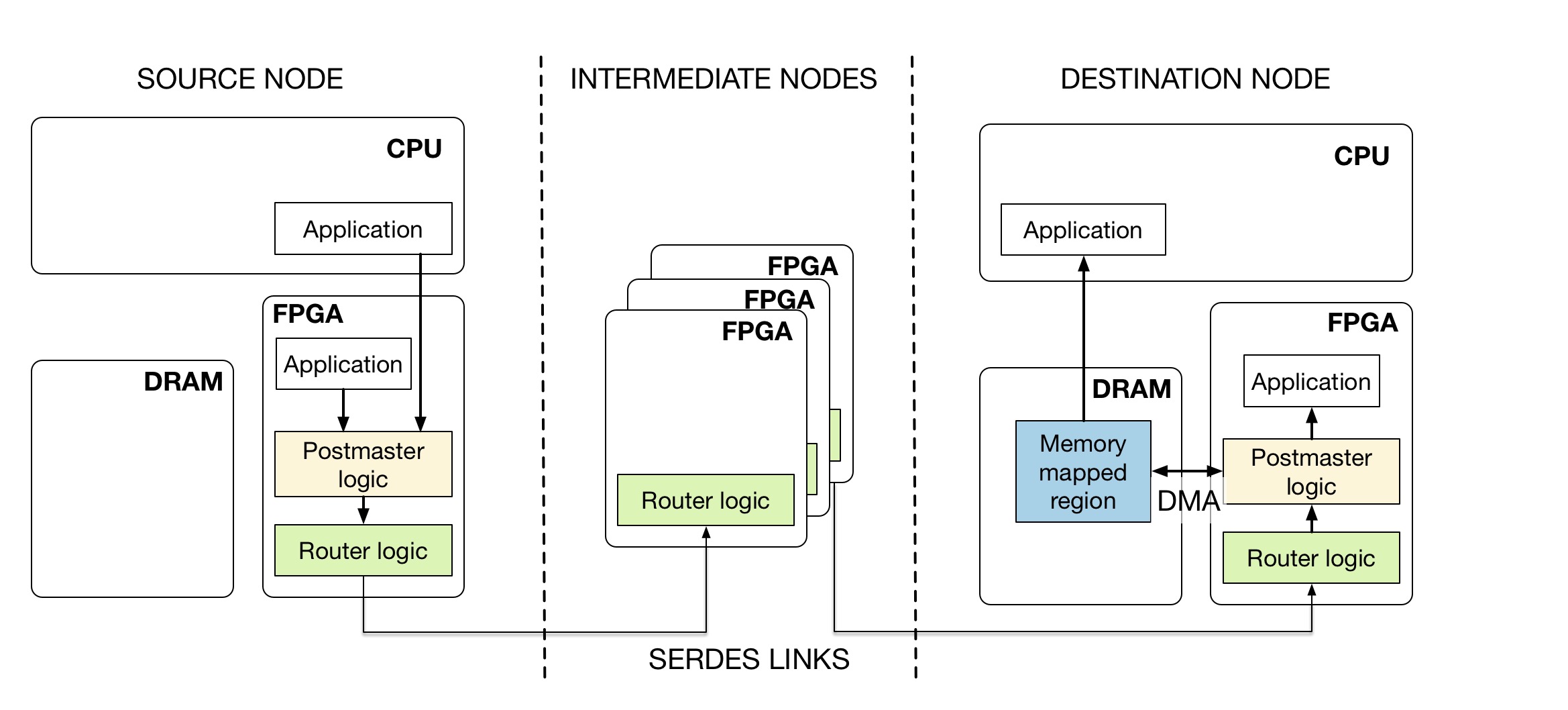}
\caption{Operation of Postmaster DMA, which is a lightweight, high bandwidth interface: An application on the source node (either on the CPU or on the FPGA) writes data to a transmit queue on the FPGA logic. This data is then received on a destination node, where it can be either consumed in the destination FPGA and/or written into a memory mapped region as shown. System software is only involved in the initialization and tear-down process, and is not shown.}
	\label{fig:post_int}
\end{figure*}

The postmaster DMA logic provides a method to move small amounts of data between nodes in the system. The function is intended to be used directly by machine intelligence application code on the software or by application hardware modules on the FPGA. It provides a communications channel with much lower overhead than going through the TCP/IP stack.

Postmaster DMA is a tunneled queue model (Fig.~\ref{fig:post_int}), where the processor (or hardware module) sees a queue that can be written at a known, fixed, address. Data written to that queue is transferred to a remote node where it is picked up by a DMA engine and moved to a pre-allocated buffer in system memory. Multiple initiators may send data to the same target. At the target, the received data is stored in a linear stream in the order in which it is received. The Postmaster hardware guarantees that a packet of data from a single initiator is always stored in contiguous locations. To reiterate, it is not necessary that the postmaster DMA be used only for processor-to-processor communication; other FPGA hardware implemented on the source/destination nodes could use it too if necessary (although in some cases this may make the `DMA' superfluous). 

Packets from multiple initiators will be interleaved within the single data queue. This model is particularly well-suited to Machine Intelligence applications in which regions or learners are distributed across multiple nodes, and each node generates multiple small outputs during each time step which become the inputs in the next time step. The function of Postmaster is to allow the node to send those outputs to their intended targets as they are generated rather than collect them and send them out as a larger transmission at the end of the time step. In addition to eliminating the burden of aggregating the data, this approach also allows much more overlap of computation and communication.

\subsection{Bridge FIFO}

\begin{figure*}[!t]
    \centering
    \includegraphics[width=\textwidth]{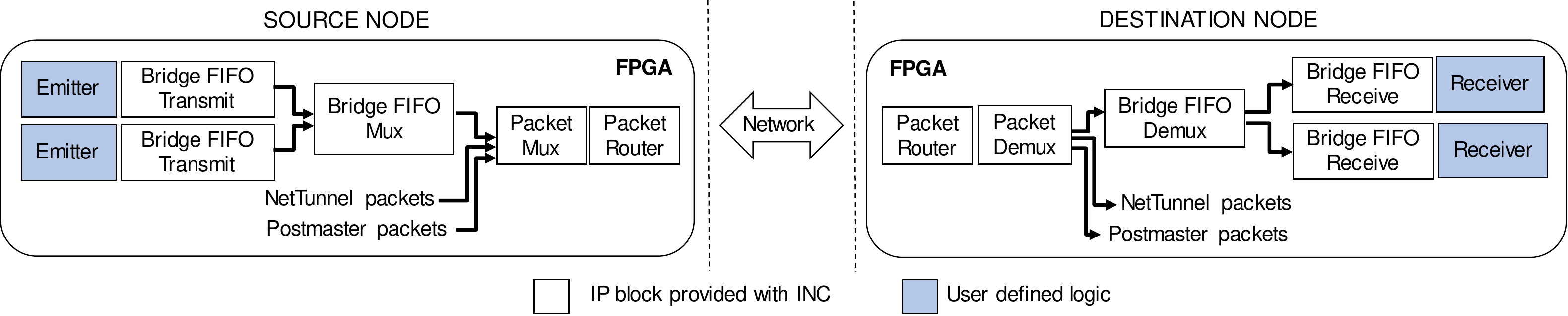}
    \caption{Implementation of a Bridge FIFO in the FPGA Logic}
    \label{fig:bridge_fifo_figure}
\end{figure*}

The bridge FIFO is intended to facilitate direct hardware to hardware communication
between two hardware modules located in separate FPGAs by exposing a regular FIFO interface. The bridge FIFO takes care of assembling the data into network packets and communicating with the packet router logic. Figure \ref{fig:bridge_fifo_figure} presents an implementation of the bridge FIFO.

The interface is composed of two modules implemented in pairs: the Bridge FIFO transmit and the Bridge FIFO receive. The first one corresponds to the write port of the FIFO while the second corresponds to the read port. They are always implemented in pairs and must be located on the source node for the transmit unit and on the destination node for the receive unit. Together, they form a communication channel.

On the source Node, the Bridge FIFO transmit converts its input (words of data) into network packets. These Bridge FIFO packets are multiplexed with other protocol packets within the Packet Mux unit which transmits all the networks packets to the Packet Router unit. The Packet Mux unit enables coexistence of multiple communication protocols.
On the destination Node, the Packet Router transmits the received network packets to the Packet Demux unit, which separates the various protocol packets and directs them to their corresponding receiver. The Bridge FIFO receive unit receives the packets and converts them back into words of data.

If multiple independent communication channels are required, then multiple pairs of Bridge FIFO transmit and receive can be instantiated. The Bridge FIFO transmits will be multiplexed within the Bridge FIFO mux and the bridge FIFO receives will be demultiplexed within the Bridge FIFO demux.
The Bridge FIFO Mux (respectively Demux) supports up to 32 Bridge FIFO transmit (respectively receive). If more channels are required, then another Bridge FIFO Mux (respectively Demux) must be instantiated.

The Bridge FIFO supports different configurable bit-widths ranging from 7 to 64. If a wider FIFO is needed, then multiple bridge FIFOs must be used in parallel to achieve the required width.

Table \ref{tab:bridge_fifo_table} presents the measured latency when using the bridge FIFO. Number of hops being zero correspond to the case where emitter and receiver are on the same node, where the latency corresponds the the delay incurred by the Bridge FIFO logic alone. The cases with 1, 3 and 6 hops are the best, average and worst case respectively on a single card system (27 nodes arranged in a cubic 3D mesh with 3 nodes on each edge).

\begin{table}[!t]
\renewcommand{\arraystretch}{1.3}
\caption{Communication latency of the bridge FIFO between two nodes.  }
\label{tab:bridge_fifo_table}
\centering
    \begin{tabular}{|c|c|c|c|c|}
         \hline
         Number of hops (in Node) & 0 & 1 & 3 & 6\\
         \hline
         Latency (in $\upmu$s) & 0.25 & 1.1 & 2.5 & 4.7\\
         \hline
    \end{tabular}
\end{table}

\section{Diagnostic Capabilities}\label{sec:diag}
INC features a wide array of diagnostic capabilities built into the hardware platform. 
This is especially important in a development platform, as the reconfigurable hardware, the system software and the application software are all concurrently evolving. While some of these have been mentioned in passing in earlier sections, we present a more detailed discussion here.  

\subsection{JTAG}
Each INC card has a single JTAG chain that is daisy chained through all 27 Zynq FPGAs. The JTAG chain can be used to access both the individual processors and the FPGAs, as these appear as different devices on chain.  Therefore, it can be used for a broad variety of tasks including configuring the FPGAs, loading code, debugging FPGA logic with Xilinx Chipscope, and debugging the ARM code through the ARM debug access port (DAP). This mode of debug is especially useful in ironing out issues during initial system bring up, when other modes may not yet be readily available. 

\subsection{Ring Bus and NetTunnel}
The Ring Bus is a sideband communications channel that links all 27 nodes on the escape card. 
The bus is implemented as a ring composed of 27 unidirectional point-to-point links. 
The topology allows data transfer between any two nodes by forwarding request and write data or read response through the intervening nodes. 
The topology also supports broadcast write operations by forwarding a given write command to all nodes on the ring. 
The routing of ring traffic is controlled by the hardware with no processor intervention.

The NetTunnel logic is functionally similar to the Ring Bus, but uses the network fabric as the transport as opposed to a dedicated sideband channel. This allows the NetTunnel logic to span the entire system, whereas the Ring Bus is confined to a single card. Note that NetTunnel does not automatically make the Ring Bus superfluous, as the network and router logic can change depending on the demands of the applications. In this scenario, having a dedicated and reliable sideband for communication is particularly useful. 

As both of these mechanisms have access to the entire 4GB address space on each node they can reach, they can be used in a wide variety of scenarios. For instance, debugging reconfigurable logic often involves reading a set of hardware registers to determine the current status, active interrupts, errors and so forth. This is especially useful when communication between nodes is involved, as the issue could be at the source, the destination or indeed along links on intermediate nodes. Similarly, checkpoints, statistics or relevant program data may be written into, and subsequently retrieved from, hardware registers. This can be very useful in debugging application or device driver code if \emph{stdout} is not available or is too late, such as in a hang scenario. 

\subsection{PCIe Sandbox}
PCIe Sandbox is an interactive utility that runs on a host x86 machine and provides access to the INC system through the PCIe interface on node (000). Using a set of simple commands, a user can read and write to addresses on all nodes in the INC system. PCIe Sandbox also supports a `read all' command that uses the Ring Bus to retrieve data from the same address location on all nodes of the card. Underneath, PCIe Sandbox `translates' these commands into read or write requests on the Ring Bus and NetTunnel mechanisms described above. This abstraction layer proves very useful for rapid debugging.

By reading and writing to registers on a node, special tasks can be accomplished including attaching the UART serial console to a particular node, reading bitstream build IDs for all the nodes, temperature of the card, EEPROM information (which may contain useful information such as USB-UART serial number, MAC ID of the gateway Ethernet interface on (100) etc), and the system configuration (i.e. how many cards are on the system).  

Finally, PCIe Sandbox is capable of loading chunks of data into the DRAM of the compute nodes. This is the preferred method for system boot up, as one can broadcast the kernel images and associated device-trees to all the nodes in the system and then write to a boot command register that initiates boot. The same method is used to configure the FPGAs in the system with new bitstreams or to program FLASH chips in the individual nodes.  

Programming the FPGAs and FLASH using the PCIe host connection and internal network is much faster than programming over JTAG.
For example, programming 27 FPGAs on a single card over JTAG takes approximately 15 minutes.  On the other hand, programming 27 FPGAs on a single card over PCIe takes a couple of seconds, including the data transfer.  Similarly, programming 432 FPGAs on 16 cards is nearly identical to programming one card, thanks to the network broadcast capability. It is important to note that JTAG can only work on a single card.  The programming speed advantage is even more pronounced for programming the FLASH chips. On one occasion, it took more than 5 hours to program 27 FLASH chips on a single card over JTAG.  In contrast, it takes about 2 minutes to program 1, 16, or 432 (or anything in between) FLASH chips over the PCIe interface.
\subsection{ARM Processor}
The ARM processor provides another dimension of per-node debug and diagnostic capability. Getting Linux running on the ARM immediately makes available a rich and varied set of diagnostic utilities. Network utilities such as \texttt{iperf} \cite{tirumala:2005}, \texttt{tcpdump} \cite{tcpdump:2018} and even \texttt{ping} \cite{muuss:2010} were used in debugging the internal Ethernet interface and the associated device driver. Counters on the ARM Performance Monitoring Unit can be readily accessed by \texttt{perf} \cite{weaver:2013} or other utilities to understand bottlenecks in application code and identify candidates for FPGA offload. Standard debuggers such as \texttt{gdb} \cite{gdb:2018} or \texttt{valgrind} \cite{valgrind:2018} could also be cross-compiled and ported, though this is untested at this time.

\section{Conclusions}\label{sec:conclusion}
We presented an overview of the INC machine intelligence platform, and publications on applications will follow shortly.  Here we focused on highlighting the main features of the hardware including flexibility, configurability, high bandwidth, highly interconnected node-to-node communication modes, and diagnostic capabilities. The uniqueness of our system is that each of the 432 nodes allows application-specific offload and this feature is not available on any other parallel machine of this scale. 


The practical challenges of using the INC systems lie at the interface of hardware and software. Programming directly at the FPGA level is often a daunting task for machine learning practitioners who are used to high-level languages and frameworks. On the other hand, hardware designers are accustomed to working with well-defined specifications and functionality, as circuit and micro-architecture trade-offs are often different as the task being offloaded change. This is a rather significant disconnect, exacerbated by the fast-paced, rapidly evolving field of artificial intelligence. 

\ifCLASSOPTIONcompsoc
  \section*{Acknowledgments}
\else
  \section*{Acknowledgment}
\fi

The authors would like to thank Spike Narayan and Jeff Welser for management support. C. Cox thanks Mike Mastro and Kevin Holland for their support in manufacturing the cards and the system.

\ifCLASSOPTIONcaptionsoff
  \newpage
\fi



\bibliographystyle{IEEEtran}
\bibliography{IEEEabrv,example_paper}

\begin{thebibliography}{10}
\providecommand{\url}[1]{#1}
\csname url@samestyle\endcsname
\providecommand{\newblock}{\relax}
\providecommand{\bibinfo}[2]{#2}
\providecommand{\BIBentrySTDinterwordspacing}{\spaceskip=0pt\relax}
\providecommand{\BIBentryALTinterwordstretchfactor}{4}
\providecommand{\BIBentryALTinterwordspacing}{\spaceskip=\fontdimen2\font plus
\BIBentryALTinterwordstretchfactor\fontdimen3\font minus
  \fontdimen4\font\relax}
\providecommand{\BIBforeignlanguage}[2]{{%
\expandafter\ifx\csname l@#1\endcsname\relax
\typeout{** WARNING: IEEEtran.bst: No hyphenation pattern has been}%
\typeout{** loaded for the language `#1'. Using the pattern for}%
\typeout{** the default language instead.}%
\else
\language=\csname l@#1\endcsname
\fi
#2}}
\providecommand{\BIBdecl}{\relax}
\BIBdecl

\bibitem{sijstermans:2018}
F.~Sijstermans, ``{The NVIDIA Deep Learning Accelerator},'' in
  \emph{Proceedings of Hot Chips 30}, Cupertino, CA, 2018.

\bibitem{dally:2018}
W.~J. Dally, C.~T. Gray, J.~Poulton, B.~Khailany, J.~Wilson, and L.~Dennison,
  ``{Hardware-Enabled Artificial Intelligence},'' in \emph{2018 Symposium on
  VLSI Circuits Digest of Technical Papers}, Honolulu, HI, 2018.

\bibitem{courbariaux:2014}
\BIBentryALTinterwordspacing
M.~Courbariaux, Y.~Bengio, and J.~David, ``Low precision arithmetic for deep
  learning,'' \emph{CoRR}, vol. abs/1412.7024, 2014. [Online]. Available:
  \url{http://arxiv.org/abs/1412.7024}
\BIBentrySTDinterwordspacing

\bibitem{gupta:2015}
\BIBentryALTinterwordspacing
S.~Gupta, A.~Agrawal, K.~Gopalakrishnan, and P.~Narayanan, ``Deep learning with
  limited numerical precision,'' in \emph{Proceedings of the 32nd International
  Conference on Machine Learning}, ser. Proceedings of Machine Learning
  Research, F.~Bach and D.~Blei, Eds., vol.~37.\hskip 1em plus 0.5em minus
  0.4em\relax Lille, France: PMLR, 07--09 Jul 2015, pp. 1737--1746. [Online].
  Available: \url{http://proceedings.mlr.press/v37/gupta15.html}
\BIBentrySTDinterwordspacing

\bibitem{han:2015}
\BIBentryALTinterwordspacing
S.~Han, H.~Mao, and W.~J. Dally, ``Deep compression: Compressing deep neural
  network with pruning, trained quantization and huffman coding,'' \emph{CoRR},
  vol. abs/1510.00149, 2015. [Online]. Available:
  \url{http://arxiv.org/abs/1510.00149}
\BIBentrySTDinterwordspacing

\bibitem{cchen:2018}
C.~Chen, J.~Choi, K.~Gopalakrishnan, V.~Srinivasan, and S.~Venkataramani,
  ``Exploiting approximate computing for deep learning acceleration,'' in
  \emph{2018 Design, Automation Test in Europe Conference Exhibition (DATE)},
  March 2018, pp. 821--826.

\bibitem{jouppi:2018}
\BIBentryALTinterwordspacing
N.~P. Jouppi, C.~Young, N.~Patil, D.~A. Patterson, G.~Agrawal, R.~Bajwa,
  S.~Bates, S.~Bhatia, N.~Boden, A.~Borchers, R.~Boyle, P.~Cantin, C.~Chao,
  C.~Clark, J.~Coriell, M.~Daley, M.~Dau, J.~Dean, B.~Gelb, T.~V. Ghaemmaghami,
  R.~Gottipati, W.~Gulland, R.~Hagmann, R.~C. Ho, D.~Hogberg, J.~Hu, R.~Hundt,
  D.~Hurt, J.~Ibarz, A.~Jaffey, A.~Jaworski, A.~Kaplan, H.~Khaitan, A.~Koch,
  N.~Kumar, S.~Lacy, J.~Laudon, J.~Law, D.~Le, C.~Leary, Z.~Liu, K.~Lucke,
  A.~Lundin, G.~MacKean, A.~Maggiore, M.~Mahony, K.~Miller, R.~Nagarajan,
  R.~Narayanaswami, R.~Ni, K.~Nix, T.~Norrie, M.~Omernick, N.~Penukonda,
  A.~Phelps, J.~Ross, A.~Salek, E.~Samadiani, C.~Severn, G.~Sizikov,
  M.~Snelham, J.~Souter, D.~Steinberg, A.~Swing, M.~Tan, G.~Thorson, B.~Tian,
  H.~Toma, E.~Tuttle, V.~Vasudevan, R.~Walter, W.~Wang, E.~Wilcox, and D.~H.
  Yoon, ``In-datacenter performance analysis of a tensor processing unit,''
  \emph{CoRR}, vol. abs/1704.04760, 2017. [Online]. Available:
  \url{http://arxiv.org/abs/1704.04760}
\BIBentrySTDinterwordspacing

\bibitem{SilverHuangEtAl16nature}
D.~Silver, A.~Huang, C.~J. Maddison, A.~Guez, L.~Sifre, G.~van~den Driessche,
  J.~Schrittwieser, I.~Antonoglou, V.~Panneershelvam, M.~Lanctot, S.~Dieleman,
  D.~Grewe, J.~Nham, N.~Kalchbrenner, I.~Sutskever, T.~Lillicrap, M.~Leach,
  K.~Kavukcuoglu, T.~Graepel, and D.~Hassabis, ``Mastering the game of {Go}
  with deep neural networks and tree search,'' \emph{Nature}, vol. 529, no.
  7587, pp. 484--489, jan 2016.

\bibitem{rajagopalan:2011}
V.~Rajagopalan, V.~Boppana, S.~Dutta, B.~Taylor, and R.~Wittig, ``Xilinx
  zynq-7000 epp: An extensible processing platform family,'' in \emph{2011 IEEE
  Hot Chips 23 Symposium (HCS)}, Aug 2011, pp. 1--24.

\bibitem{crockett:2014}
L.~H. Crockett, R.~A. Elliot, M.~A. Enderwitz, and R.~W. Stewart, \emph{The
  Zynq Book: Embedded Processing with the Arm Cortex-A9 on the Xilinx Zynq-7000
  All Programmable Soc}.\hskip 1em plus 0.5em minus 0.4em\relax UK: Strathclyde
  Academic Media, 2014.

\bibitem{zynq7000}
\BIBentryALTinterwordspacing
Xilinx, ``Zynq 7000 technical reference manual (ug 585),'' 2018. [Online].
  Available:
  \url{https://www.xilinx.com/support/.../user_guides/ug585-Zynq-7000-TRM.pdf}
\BIBentrySTDinterwordspacing

\bibitem{furber:2014}
S.~B. Furber, F.~Galluppi, S.~Temple, and L.~A. Plana, ``The spinnaker
  project,'' \emph{Proceedings of the IEEE}, vol. 102, no.~5, pp. 652--665, May
  2014.

\bibitem{painkras:2013}
E.~Painkras, L.~A. Plana, J.~Garside, S.~Temple, F.~Galluppi, C.~Patterson,
  D.~R. Lester, A.~D. Brown, and S.~B. Furber, ``Spinnaker: A 1-w 18-core
  system-on-chip for massively-parallel neural network simulation,'' \emph{IEEE
  Journal of Solid-State Circuits}, vol.~48, no.~8, pp. 1943--1953, Aug 2013.

\bibitem{krupnova:2004}
H.~Krupnova, ``Mapping multi-million gate socs on fpgas: industrial methodology
  and experience,'' in \emph{Proceedings Design, Automation and Test in Europe
  Conference and Exhibition}, vol.~2, Feb 2004, pp. 1236--1241 Vol.2.

\bibitem{assad:2012}
\BIBentryALTinterwordspacing
S.~Asaad, R.~Bellofatto, B.~Brezzo, C.~Haymes, M.~Kapur, B.~Parker, T.~Roewer,
  P.~Saha, T.~Takken, and J.~Tierno, ``A cycle-accurate, cycle-reproducible
  multi-fpga system for accelerating multi-core processor simulation,'' in
  \emph{Proceedings of the ACM/SIGDA International Symposium on Field
  Programmable Gate Arrays}, ser. FPGA '12.\hskip 1em plus 0.5em minus
  0.4em\relax New York, NY, USA: ACM, 2012, pp. 153--162. [Online]. Available:
  \url{http://doi.acm.org/10.1145/2145694.2145720}
\BIBentrySTDinterwordspacing

\bibitem{haring:2012}
R.~Haring, M.~Ohmacht, T.~Fox, M.~Gschwind, D.~Satterfield, K.~Sugavanam,
  P.~Coteus, P.~Heidelberger, M.~Blumrich, R.~Wisniewski, a.~gara, G.~Chiu,
  P.~Boyle, N.~Chist, and C.~Kim, ``The ibm blue gene/q compute chip,''
  \emph{IEEE Micro}, vol.~32, no.~2, pp. 48--60, March 2012.

\bibitem{geist:1996}
A.~Geist, W.~Gropp, S.~Huss-Lederman, A.~Lumsdaine, E.~Lusk, W.~Saphir,
  T.~Skjellum, and M.~Snir, ``Mpi-2: Extending the message-passing interface,''
  in \emph{Euro-Par'96 Parallel Processing}, L.~Boug{\'e}, P.~Fraigniaud,
  A.~Mignotte, and Y.~Robert, Eds.\hskip 1em plus 0.5em minus 0.4em\relax
  Berlin, Heidelberg: Springer Berlin Heidelberg, 1996, pp. 128--135.

\bibitem{tirumala:2005}
\BIBentryALTinterwordspacing
A.~Tirumala, F.~Qin, J.~Dugan, J.~Ferguson, and K.~Gibbs, ``{iPerf: TCP/UDP
  bandwidth measurement tool},'' 2005. [Online]. Available:
  \url{https://iperf.fr/}
\BIBentrySTDinterwordspacing

\bibitem{tcpdump:2018}
\BIBentryALTinterwordspacing
tcpdump, ``tcpdump,'' 2018. [Online]. Available: \url{http://www.tcpdump.org/}
\BIBentrySTDinterwordspacing

\bibitem{muuss:2010}
\BIBentryALTinterwordspacing
M.~Muuss, ``{The Story of the PING Program},'' 2010. [Online]. Available:
  \url{https://www.webcitation.org/5saCKBpgH}
\BIBentrySTDinterwordspacing

\bibitem{weaver:2013}
V.~Weaver, ``Linux perf event features and overhead,'' in \emph{The 2nd
  International Workshop on Performance Analysis of Workload Optimized Systems,
  FastPath}, 01 2013.

\bibitem{gdb:2018}
\BIBentryALTinterwordspacing
gdb, ``{GDB: The GNU Project Debugger},'' 2018. [Online]. Available:
  \url{http://www.tcpdump.org/}
\BIBentrySTDinterwordspacing

\bibitem{valgrind:2018}
\BIBentryALTinterwordspacing
Valgrind, ``{Valgrind},'' 2018. [Online]. Available: \url{http://valgrind.org/}
\BIBentrySTDinterwordspacing

\end{thebibliography}
%



%
\begin{IEEEbiography}[{\includegraphics[width=1in,height=1.25in,clip=true,keepaspectratio]{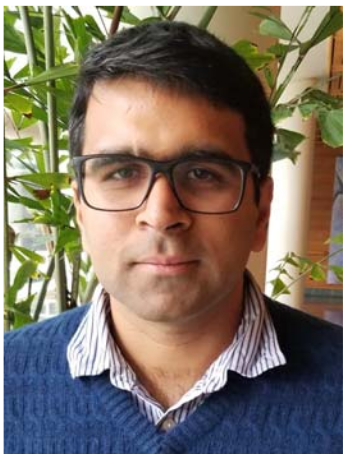}}]{Pritish Narayanan}
received    his    PhD    in    Electrical    and    Computer    Engineering from the University of Massachusetts Amherst. He joined IBM Research  –  Almaden  as  a  Research  Staff  Member  in  2013  as  part  of  the  Storage  Class  Memory/MIEC  project,  where  he  investigated  circuit  design  challenges  for  access  devices  used  in  3D  crosspoint  memory.  His  current  research  interests  are  in  the  area  of  ultra-high-performance  hardware  systems  for  Artificial  Intelligence  and  Cognitive  computing  including  i)  Novel  non-Von  Neumann  architectures  based  on  emerging  memory,  where  he  is  the  lead  circuit  architect  for  two  deep  learning  test  sites  based  on  Phase   Change   Memory   (PCM)   and   mixed-signal   hardware   and   ii)   FPGA-based  systems  exploiting  massive  parallelism  and/or  approximate  computing  techniques.  Dr.  Narayanan  has  presented  one  prior  keynote  (International  Memory  Workshop  2017)  and  a  tutorial  session  (Device  Research Conference 2017), in addition to several invited talks. He won Best Paper Awards at IEEE Computer Society Symposium on VLSI 2008 and at IEEE  Nanoarch  2013.  He  has  also  been  a  Guest  Editor  for  the  Journal  of  Emerging   Technologies   in   Computing,   the   Program   Chair   at   IEEE   Nanoarch  2015,  Special  Session  Chair  for  IEEE  Nano  2016  and  served  on  the Technical Program Committees of several conferences..
\end{IEEEbiography}

\begin{IEEEbiography}[{\includegraphics[width=1in,height=1.25in,clip,keepaspectratio]{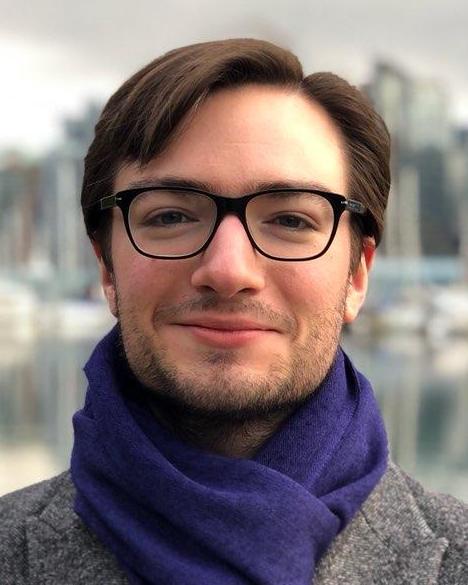}}]{Alexis Asseman}
received his M.S. degree in electrical engineering jointly from the Grenoble Institute of Technology in France, the Politecnico of Turin in Italy, and EPFL in Switzerland in 2016. He joined the IBM Almaden Research Center in San Jose, CA in 2017 as a Research Engineer focusing on Machine Learning and high-performance and/or energy efficient implementations on FPGA hardware.
\end{IEEEbiography}


\begin{IEEEbiography}[{\includegraphics[width=1in,height=1.25in,trim=.15in 0 .15in 0,clip,keepaspectratio]{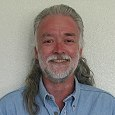}}]{Charles Cox}
received the B.S. degree in computer engineering from the University  of Illinois, Urbana, in 1983. He joined the then IBM San Jose Research Laboratory in 1984 and is currently  an  Advisory Engineer at the IBM Almaden  Research Center, San Jose, CA. He is interested in the rapid prototyping of custom hardware to verify new algorithms amenable to hardware  realizations,  and has worked in the  areas of image processing, communications, pattern recognition, and error correction.
\end{IEEEbiography}

\begin{IEEEbiography}[{\includegraphics[width=1in,height=1.25in,trim=.8in 0 .8in 0,clip,keepaspectratio]{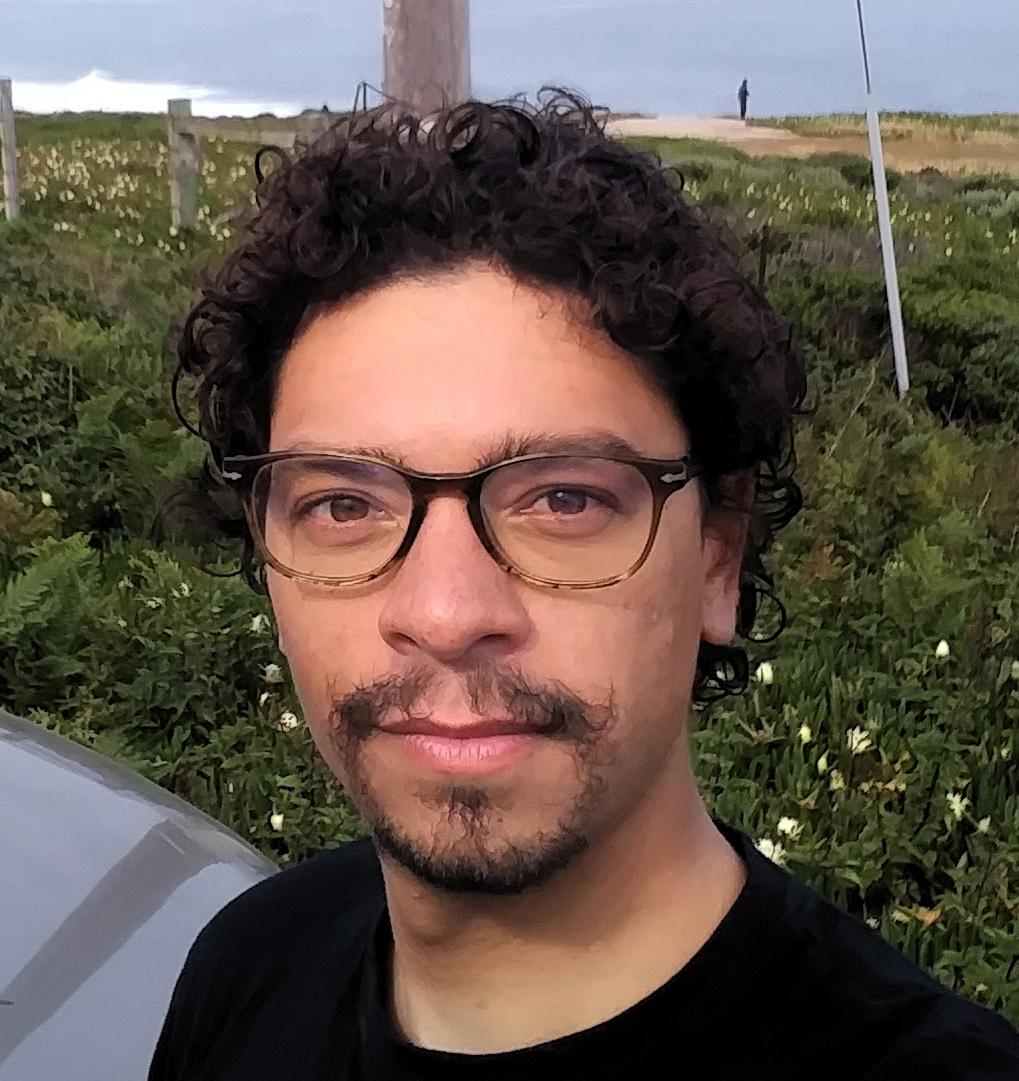}}]{Nicolas Antoine} received his M.S. degree in electrical engineering jointly from the Ecole Centrale de Nantes in France and the RWTH Aachen in Germany in 2018. He joined the IBM Almaden Research Center in San Jose, CA in 2018 as a Research Engineer working on hardware efficient implementation of deep learning algorithms on FPGA. 
\end{IEEEbiography}

\begin{IEEEbiography}[{\includegraphics[width=1in,height=1.25in,trim=.15in 0 .15in 0,clip,keepaspectratio]{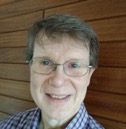}}]{Harald Huels} received the M.S degree in communications engineering from the Technische Hochschule Mittelhessen, Germany
in 1984. He worked for several companies as well as the University
of Mainz in the field of measurement and control technology as 
development engineer. After this he was self-employed and 
developed control systems based on microprocessors for industrial 
automation. He joined IBM Storage Division in 1990. Until 1998 he
worked at the Analysis Lab responsible for automation and testing,
designing complex test systems for quality assurance in the disk drive
production. From 1998 until 2001 he was on a working assignment at
IBM San Jose, CA responsible for development and research of new
test methodologies focused on future products of IBM’s disk drives.
Returning from the US he joined Engineering and Technology Services
of IBM and started a research project with IBM Almaden. He was
responsible for second level packaging of the ‘Ice-Cube’ project.
In 2005 he joined IBM Research and Development, Boeblingen, Germany.
Since then he is working as Senior Hardware Architect for both
System-p and System-z with a strong focus on packaging. His major interest
is in system design utilizing the newest technologies to build servers.     
\end{IEEEbiography}

\begin{IEEEbiography}[{\includegraphics[width=1in,height=1.25in,clip,keepaspectratio]{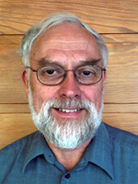}}]{Winfried Wilcke}
received a Ph.D. in experimental nuclear physics (1976) and worked at the University of Rochester, Lawrence Berkeley Lab and Los Alamos on heavy-ion nuclear reactions and meson physics. In 1983 he joined IBM Research in Yorktown Heights,NY, where he played a key role in creating IBM's line of very fast distributed memory supercomputers , including the Victor (research) and the IBM SP series of supercomputers.
In 1991, Wilcke co-founded HAL Computer Systems, where he was first Director of Architecture and later CTO. HAL grew to 450 people before it was purchased by Fujitsu. It was instrumental in creating jointly with Sun Microsystems the 64-bit SPARC architecture, which is now underlying all Sun and Fujitsu Sparc systems.
In 1996, Wilcke retired temporarily from R\&D and embarked on a long tropical sailing/diving voyage, then rejoined IBM Research in California, where, in 2001, he founded an IBM spin-out company. Later he became senior manager of Nanoscale Science \& Technology, which includes projects in quantum physics, batteries and machine learning and especially general artificial intelligence. Wilcke is co-author of over 120 publications (3500 citations) in nuclear physics, computer architecture and energy, a book  (Random Walk at Amazon) and numerous patents in computer engineering. Wilcke has been chair of several conferences, including Compcon, Hot Chips and is the founder of the ‘Beyond Lithium Ion’ conference series.
\end{IEEEbiography}

\begin{IEEEbiography}[{\includegraphics[width=1in,height=1.25in,clip,keepaspectratio]{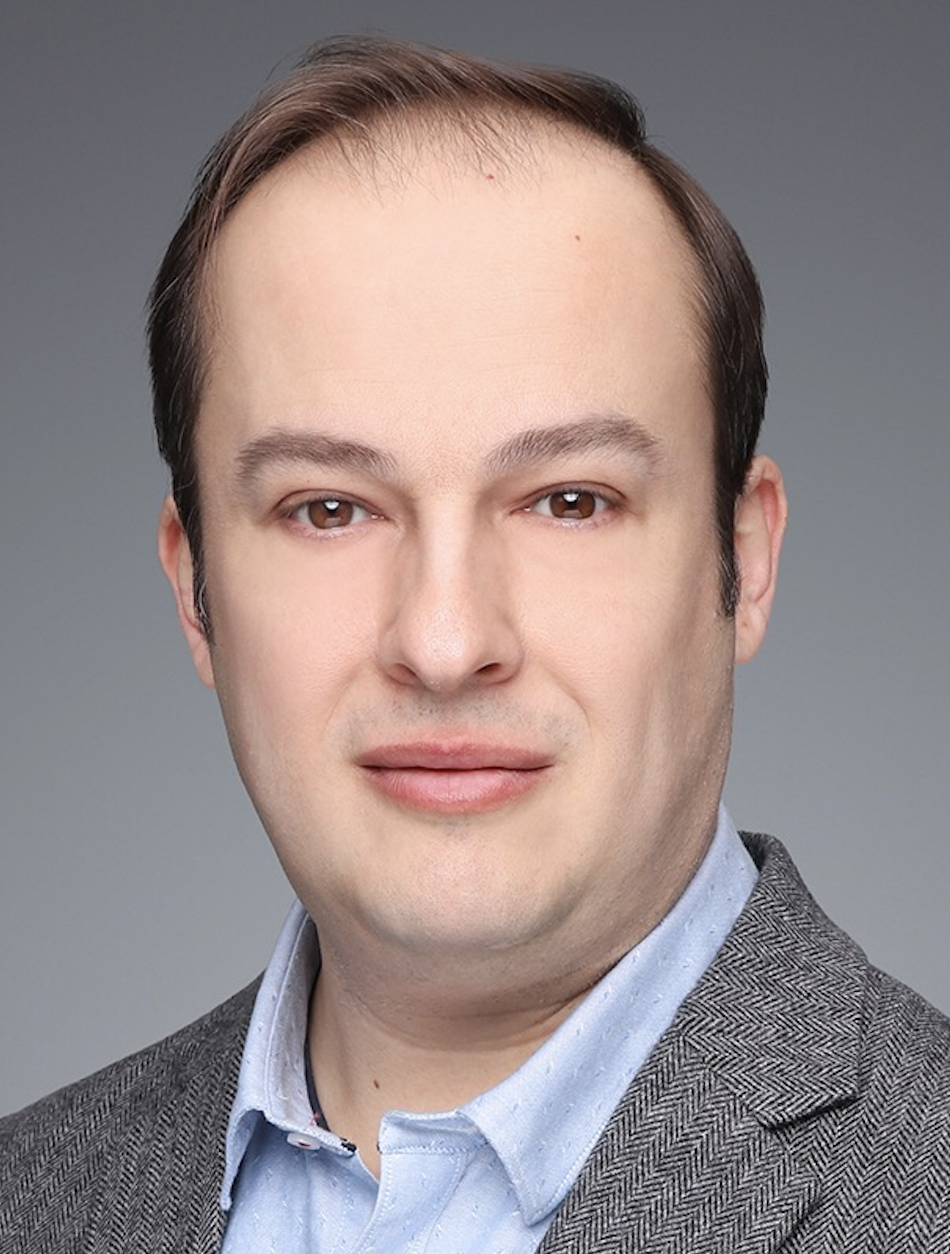}}]{Ahmet S. Ozcan}
received his Ph.D. in Physics from Boston University (2003) and joined IBM in 2006 after a postdoctoral appointment. He had worked in IBM’s Microelectronics division and contributed to the development of several CMOS technology nodes, which were commercialized in IBM’s server products and in mobile devices through industry partnerships. In 2015, after an assignment in France on FDSOI technology development, Ahmet moved to Silicon Valley to work on brain-inspired computing and artificial intelligence. He and his team currently work on visually-grounded language and reasoning using memory-augmented neural networks, as well as developing applications for FPGA based re-configurable hardware systems.  An IBM Master Inventor, Ahmet holds over 56 patents and 85 pending patent applications. He has authored and co-authored more than 50 peer-reviewed articles and one book chapter.
\end{IEEEbiography}



\end{document}